\documentclass[preprint,preprintnumbers,amsmath,amssymb]{revtex4}
\usepackage{graphicx}
\usepackage{epsfig}
\usepackage{dcolumn}
\usepackage{bm}
\begin{document}
\title{Calculation of van der Walls coefficients of alkali metal clusters by hydrodynamic
approach to time-dependent density-functional theory}
\author{Arup Banerjee$^{a}$  and Manoj K. Harbola$^{b}$\\
(a) Laser Physics Division, Centre for Advanced Technology\\
Indore 452013, India\\
(b) Department of Physics, Indian Institute of Technology, Kanpur\\
U.P 208016, India}
\begin{abstract}
In this paper we employ the hydrodynamic formulation of time-dependent density functional theory to obtain the van der Waal coefficients $C_{6}$ and $C_{8}$ of alkali-metal clusters of various sizes including very large clusters. Such calculation becomes computationally very demanding in the orbital-based Kohn-Sham formalism, but quite simple in the hydrodynamic approach. We show that for interactions between the clusters of same sizes, $C_{6}$ and $C_{8}$ sale as the sixth and the eighth power of the cluster radius rsepectively, and approach the respective classically predicted values for the large size clusters.
\end{abstract} 
\maketitle
\section{Introduction}
The long-range van der Waals forces play an important role in the
description of many physical and chemical phenomena such as
adhesion, surface tension, physical adsorption, etc. The
correlations between electron density fluctuations at widely
separated locations give rise to these long range forces. For
clusters the  knowledge of coefficients of van der Waals
interaction is useful in describing the cluster-cluster collisions
and also for characterizing the orientation of clusters in bulk
matter. The van der Waals interaction coefficients, namely
$C^{AB}_{6}$ and $C^{AB}_{8}$ between small alkali metal particles
within the spherical  jellium background (SJBM) model of  have
been theoretically calculated in the past
\cite{pacheco1,pacheco2,banerjee1} using time dependent density
functional theory (TDDFT) \cite{gross}. In Refs.
\cite{pacheco1,pacheco2} time-dependent Kohn-Sham (TDKS)
\cite{gross} formalism of TDDFT  was employed to obtain
$C^{AB}_{6}$ and $C^{AB}_{8}$. On the other hand, in Ref. \cite{banerjee1}, we employed density based method within TDDFT formalism to obtain the van der Waals
coefficient $C^{AB}_{6}$ for the clusters. The TDKS formalism is an orbital
based theory and involves calculation of these orbitals in a
self-consistent manner. Thus, as the size of system increases
these calculations tend to become computationally cumbersome. In
such cases, our approach \cite{banerjee1} is much easier to as it makes calculations size independent with hardly any loss in
accuracy. 
 
The main aim of this paper is to extend our previous study \cite{banerjee1,banerjee2,manoj1} by applying the density based hydrodynamic approach to calculate  the
higher-order van der Waals coefficient $C^{AB}_{8}$ of alkali atom
clusters characterizing dipole-quadrupole interaction of
long-range force and study its evolution with the size of
clusters. In addition to this we also calculate $C^{AB}_{6}$ and
$C^{AB}_{8}$ coefficients for the pair interaction between
potassium clusters of various sizes and between potassium and
sodium clusters of different sizes as well.

Before proceeding further, it is necessary to note that the
density functional theory (DFT) in principle should give the exact
ground-state properties including long range van der Waals energies.
However, the frequently used local density approximation (LDA)
\cite{parr,dreizler} and generalized gradient approximations (GGA)
\cite{gga1,gga2,gga3} fail to reproduce the van der Waals
energies. This is due to the fact that LDA and GGA cannot
completely simulate the correlated motion of electrons arising
from coulomb interaction between distant non overlapping
electronic systems. It is only recently that several attempts
\cite{andersson,dobson,kohn} have been made to obtain van der
Waals energies directly from the ground-state energy functional by
correcting the long range nature of the effective Kohn-Sham
potential. On the other hand, it is possible to make reliable
estimates of the van der Waals coefficients by using expressions
which relate these coefficients to the frequency dependent
multipole polarizabilities at imaginary frequencies
\cite{kaplan,stone}. We follow this route for the calculation of these
coefficients.

The paper is organized as follows: In section II we express 
$C_{6}$ and $C_{8}$ in terms the dipole and the quadrupole dynamic polarizabilities. We then briefly describe our method of obtaining these polarizabilities employing hydrodynamic approach of TDDFT. Results of our calculations are presented and discussed in section III.

\section{Methods of Calculation}
The nonretarded electrostatic interaction energy between two
spherically symmetric electronic systems ${\it A}$ and ${\it B}$
separaterd by interaction distance ${\it R}$ can be written as
\cite{kaplan,stone}
\begin{equation}
V_{AB}(R) = -\frac{C_{6}^{AB}}{R^{6}} - \frac{C_{8}^{AB}}{R^{8}} -
\cdots
\label{eq1}
\end{equation}
The above expression has been obtained by assuming that ${\it R}$
is very large so that the charge distributions corresponding to
the two systems do not overlap. The coefficient $C_{6}^{AB}$
describes the dipole-dipole interaction, and $C_{8}^{AB}$
corresponds to the dipole-quadrupole interaction between system
${\it A}$ and ${\it B}$. These coefficients can be expressed in
terms of the dynamic multipole polarizability $\alpha_{l}
(\omega)$ (where $l$ denotes the index of multipolarity) by
following relations  \cite{dalgarno,buckingham}:
\begin{eqnarray}
C_{6}^{AB} & = & C(A,1;B,1) \nonumber \\
C_{8}^{AB} & = & C(A,1;B,2) + C(A,2;B,1)
 \label{eq2}
\end{eqnarray}
with
\begin{equation}
C(A,l_{1};B,l_{2}) = \frac{\left ( 2l_{1} + 2l_{2}\right
)!}{2\pi(2l_{1})!(2l_{2})!}\int_{0}^{\infty}d\omega\alpha_{l_{1}}^{A}(i\omega
)\alpha_{l_{2}}^{B}(i\omega ) \label{eq3}
\end{equation}
where $\alpha_{l}^{X}(i\omega )$ is the multipole polarizability
of system  $X$ ($ X= A$ or $B$) at imaginary frequency $u =
i\omega$. Although $\alpha_{l}^{X}(i\omega )$ does not have any
physical significance, expressions given by Eqs. (\ref{eq2}) and (\ref{eq3}) make the computation  of $C_{6}^{AB}$ and $C_{8}^{AB}$ straightforward. 
Moreover, mathematically $\alpha_{l}^{X}(i\omega
)$ is better behaved than its real frequency counterpart $\alpha (\omega )$:
it does not have any singularity and decreases monotonically from its static value
$\alpha_{l}^{X}(0)$ to zero as $\omega\rightarrow\infty$.
Consequently, the quadrature in Eq. (\ref{eq3}) can be computed
quite accurately. To determine the frequency dependent polarizabilities appearing
in Eq.(\ref{eq3}), we employ variation-perturbation method within the hydrodynamic approach of TDDFT. We now describe the theory in brief. For details the reader is referred to the literature \cite{banerjee1,banerjee2,manoj1}

The basic dynamical variables of the hydrodynamic theory are the
time dependent density $\rho ({\bf r},t)$ and the velocity
potential $S({\bf r},t)$. Thus the total time-averaged energy can be
expressed in terms of these two variables. For our purpose we need to evaluate
the second-order change in the time-averaged energy as this is
directly related to the frequency dependent multipole
polarizability by the relation
\begin{equation}
\alpha_{l}(\omega ) = -4E_{l}^{(2)} \label{eq12}
\end{equation}
The second-order time-averaged energy $E_{l}^{(2)}$ in turn can be
expressed as
\begin{eqnarray}
E_{l}^{(2)}& =& \left\{\frac{1}{2}\int\frac{\delta^{2}
F[\rho]}{\delta\rho({\bf r},t)\delta\rho({\bf r'},t)}
\rho^{(1)}({\bf r},t)\rho^{(1)}({\bf r'},t)d{\bf r}d{\bf r'}
+ \int v_{app}^{(l)}({\bf r},t)\rho^{(0)}({\bf r})d{\bf r}\right. \nonumber \\
& + & \left.\int\frac{\partial S^{(1)}({\bf r},t)}{\partial
t}\rho^{(1)}({\bf r},t)d{\bf r} + \frac{1}{2}\int
({\bf\nabla}S^{(1)}\cdot{\bf\nabla}S^{(1)}) \rho^{(0)}({\bf
r})d{\bf r}\right\}, \label{eq13}
\end{eqnarray}
where the curly bracket denotes the time averaging over a period
of the applied oscillating field and $\rho^{(0)}({\bf r})$
represents the ground-state density. It is easily shown \cite{banerjee1} that
$E_{l}^{(2)}$ is stationary with respect to the variations in the 
first-order induced density $\rho^{(1)}({\bf r},t)$ and the
induced current-density $S^{(1)}({\bf r},t)$. Consequently,
$E_{l}^{(2)}$ can be determined by choosing appropriate
variational forms for $\rho^{(1)}({\bf r},t)$ and $S^{(1)}({\bf
r},t)$ and making $E_{l}^{(2)}$ stationary with respect to the variations in the 
parameters of $\rho^{(1)}({\bf r},t)$ and $S^{(1)}({\bf r},t)$. In
the above expression the functional 
\begin{equation}
F[\rho ] = T_{s}[\rho ] + E_{H}[\rho ] + E_{XC}[\rho ],  \label{eq14}
\end{equation}
where $T_{s}[\rho ]$, $E_{H}[\rho ]$  and $E_{XC}[\rho ]$ denote the kinetic, Hartree and 
the exchange-correlation (XC) energy functionals respectively. The
exact forms of $T_{s}[\rho ]$ and $E_{XC}[\rho ]$ are not known. Consequently to
perform any calculation one needs to use approximate forms for
these functionals. On the other hand, the hartree energy
functional $E_{H}[\rho ]$ representing classical coulomb energy is
exactly known and it given by
\begin{equation}
E_{H}[\rho ] = \frac{1}{2}\int\int\frac{\rho ({\bf r},t)\rho ({\bf
r'},t)}{|{\bf r} - {\bf r'}|}d{\bf r}d{\bf r'}. \label{hartree}
\end{equation}
For the purpose of calculation the multipolar applied potential
$v_{app}^{l}({\bf r},t)$ is chosen to be
\begin{equation}
v_{app}^{l}({\bf r},t) = {\cal E}r^{l}Y_{l0}(\theta,\phi)cos\omega
t \label{eq15}
\end{equation}
where ${\cal E}$ and $\omega$ represent the amplitude and the
frequency of the applied periodic electromagnetic field. In
accordance with the above form of the applied potential the
variational forms for  $\rho^{(1)}({\bf r},t)$ and $S^{(1)}({\bf
r},t)$ are chosen to be
\begin{eqnarray}
\rho^{(1)}({\bf r},t) & = & \rho^{(1)}({\bf r},\omega)cos\omega t \nonumber \\
S^{(1)}({\bf r},t) & = & \omega S^{(1)}({\bf r},\omega)sin\omega t
\label{14a}
\end{eqnarray}
with
\begin{eqnarray}
\rho^{(1)}({\bf r},\omega) & = & \sum_{i}c_{i}r^{i}\rho^{(0)}({\bf
r})Y_{l0}(\theta,\phi)
\nonumber \\
S^{(1)}({\bf r},\omega) & = & \sum_{i}d_{i}r^{i}({\bf
r})Y_{l0}(\theta,\phi)\label{14b}
\end{eqnarray}
where $\rho^{(0)}(\bf r)$ is the ground-state density and $c_{i}$
and $d_{i}$ are the variational parameters obtained by minimizing
time-averaged second-order energy $E_{l}^{(2)}$. On substituting
Eq. (\ref{14a}) in Eq. (\ref{eq13}) and taking average over time
we get
\begin{eqnarray}
E^{(2)}& =& \frac{1}{4}\int\frac{\delta^{2}
F[\rho]}{\delta\rho({\bf r})\delta\rho({\bf r'})} \rho^{(1)}({\bf
r},\omega)\rho^{(1)}({\bf r'},\omega)d{\bf r}d{\bf r'} +
\frac{1}{2}\int v_{app}^{(1)}({\bf r})\rho^{(1)}({\bf r},\omega)
d{\bf r} \nonumber \\
& + & \frac{\omega^{2}}{2}\int S^{(1)}({\bf r},\omega)
\rho^{(1)}({\bf r},\omega)d{\bf r} + \frac{\omega^{2}}{4}\int
({\bf\nabla}S^{(1)}\cdot{\bf\nabla}S^{(1)}) \rho^{(0)}({\bf
r})d{\bf r}, \label{11}
\end{eqnarray}
At this point it is important to point out that the VP method
discussed above is also applicable for the imaginary frequencies
with $\omega^{2}$ replaced by $-\omega^{2}$ in Eq. (\ref{11})
\cite{banerjee1}. This allows us to
determine dynamic multipolarizability at imaginary frequencies
($\alpha (i\omega)$) by exactly the same procedure as employed for
getting $\alpha (\omega)$. All that is required for this is to
change $\omega^{2}$ to $-\omega^{2}$ in Eq. (\ref{11}). This is
done very easily in the numerical code written for determining
dynamic polarizability at real frequencies.

As mentioned earlier the calculation of $E_{l}^{(2)}$ requires
approximating the functionals $T_{s} [\rho ]$ and $E_{XC}[\rho ]$. To
this end we choose the von Weizsacker \cite{weizsacker} form for 
$T_{s} [\rho ]$ which is given as
\begin{equation}
T_{W}[\rho] =
\frac{1}{8}\int\frac{{\bf\nabla}\rho\cdot{\bf\nabla}\rho}
{\rho}d{\bf r}. \label{14}
\end{equation}
Our previous experience with the calculation of response
properties has led us to choose
von Weizsacker KE functional for the polarizability calculation.
For the XC energy, adiabatic local-density approximation (ALDA)
\cite{gross} is accurate enough to describe the energy changes.
Thus the exchange energy is approximated by the Dirac exchange
functional \cite{dirac}
\begin{eqnarray}
E_{x}[\rho] & = & C_{x}\int\rho^{\frac{4}{3}}({\bf r})d{\bf r}\nonumber \\
C_{x} & = & -\frac{3}{4}\left (\frac{3}{\pi}\right
)^{\frac{1}{3}}. \label{15}
\end{eqnarray}
and for the correlation energy functional we employ the
Gunnarsson-Lundqvist (GL) \cite{gunnarsson} parametrized form for
it .

In the present paper the ground-state densities $\rho^{(0)}({\bf
r})$ of clusters are
obtained by employing purely density-based extended Thomas-Fermi
(ETF) \cite{brack2,brack} method within the spherical jellium background model (SJBM) of metal clusters.
This approach yields the ground-state densities of very large
clusters (containing up to 1000 atoms) easily thereby allowing us
to study the evolution of van der Waals coefficients with the size
of clusters. For details of the ETF method and its application to
study alkali-metal cluster, we refer the reader to
\cite{brack2,brack,brack3,banerjee2,manoj1}. In the next section we
present the results for $C^{AB}_{6}$ and $C^{AB}_{8}$ of alkali
metal clusters by employing the method describe above.

\section{Results and Discussion}
We have performed calculations for the coefficients $C^{AB}_{6}$ and
$C^{AB}_{8}$  between clusters of alkali metal atoms of various
sizes. In this paper we consider clusters made up of sodium
($r_{s} = 4.0 a.u.$) and potassium ($r_{s} = 4.86 a.u.$) atoms (
where $r_{s}$ is the Wigner-Seitz radius of the cluster species).
First we discuss the results for the sodium clusters. For
completeness we also include $C^{AB}_{6}$ numbers of sodium
clusters in this paper. As in our previous study, Table I and II we present the results
for $C^{AB}_{6}$ and $C^{AB}_{8}$ respectively, between sodium
clusters containing 2, 8, 20, and 40 atoms. For comparison we also
show the corresponding TDKS results of Ref. \cite{pacheco2} in
parenthesis. Table I and II clearly show that the numbers obtained
by employing the hydrodynamic approach for small clusters are
quite close to the corresponding numbers obtained by the TDKS
method. The numbers for $C^{AB}_{6}$ obtained in this paper are
systematically lower than the orbital-based TDKS results. The
difference between the two results is slightly more for larger
clusters than the smaller ones, the maximum difference being of
the order of 10$\%$.  Now we discuss the main results of this
paper, that is, numbers for $C^{AB}_{8}$ between clusters of
sodium clusters.

In Table II we present the results for the coefficient $C^{AB}_{8}$. 
This table clearly shows that our numbers for $C^{AB}_{8}$ are quite close to the
corresponding TDKS results. In comparison to the results for
$C^{AB}_{6}$ our numbers for $C^{AB}_{8}$ are closer to the
corresponding TDKS values.  Moreover, unlike $C^{AB}_{6}$
case, the numbers for $C^{AB}_{8}$ are not always lower than the
corresponding TDKS results.  The values of $C^{AB}_{8}$ between 2
and 2 and, 2 and 8 atom sodium clusters obtained by the hydrodynamic
method are slightly more than the corresponding TDKS results.
We next apply hydrodynamic approach to calculate the van der Waals
coefficients between potassium clusters of various sizes. 

In Table III and IV we present the results for $C^{AB}_{6}$ and
$C^{AB}_{8}$ respectively, between potassium clusters of different
sizes. Again two tables clearly show that both $C^{AB}_{6}$ and
$C^{AB}_{8}$ between potassium clusters are obtained quite
accurately with the hydrodynamic approach. The numbers obtained by
us for potassium clusters are all lower than the corresponding
TDKS results. Table III and IV also clearly show that the
difference between the hydrodynamic and the TDKS results is lower
for the larger clusters than than the smaller ones. On the other
hand, for sodium clusters the difference between the two results shows the opposite
trend.

Next we now present in Table V and VI the results for
$C^{AB}_{6}$ and $C^{AB}_{8}$ respectively, for the pair
interaction between sodium and potassium clusters with 2, 8, and,
20 constituent atoms. Once again we see from these two table
that hydrodynamic numbers quite close to the corresponding TDKS
results barring few exceptions like the number for $C^{AB}_{6}$
between two types of clusters containing 8 atoms each. Table V and
VI also show that like the pair interaction between clusters of
identical atoms the value of $C^{AB}_{8}$ for two different
clusters obtained by hydrodynamic approach is closer to the TDKS
as compared to the corresponding $C^{AB}_{6}$ results.

With favourable comparison between the numbers obtained by our approach and those by the TDKS approach for small clusters we now employ the hydrodynamic approach to evaluate the van der Waal coefficients for much larger clusters. This allows us to study 
the size dependence of van
der Waals coefficients and how they approach the asymptotic classical
values. In the present paper we have performed calculations for clusters
containing up to 1000 atoms without any increase in computational
effort. To study the size dependence of van der Waals coefficients
we plot in Figs. 1 and 2, $C^{AB}_{6}/R_{0}^{6}$ and
$C^{AB}_{8}/R_{0}^{8}$ respectively, as functions of  $R_{0}$ (
where $R_{0} = r_{s}N^{1/3}$ and N denotes number of atoms). These
two figures clearly exhibit that as the size of cluster increases
the values of coefficients $C^{AA}_{6}/R_{0}^{6}$ and
$C^{AA}_{8}/R_{0}^{8}$ saturate to constants numbers, indicating
that $C^{AA}_{6}$ and $C^{AA}_{8}$ scale as the sixth and the
eight power of the radius $R_{0}$ of the cluster respectively.
This trend is consistent with the fact that the properties of
metal clusters approach their corresponding classical values as
the size of cluster is increased. The classical expressions for
$C^{AA}_{6}/R_{0}^{6}$ and $C^{AA}_{8}/R_{0}^{8}$ between clusters
of same sizes  can be written as \cite {pacheco2}
\begin{eqnarray}
C^{AA}_{6}/R_{0}^{6} & = & \frac{3}{4}\omega_{Mie} \nonumber \\
C^{AA}_{8}/R_{0}^{8} & = & \frac{15}{2}\frac{\sqrt{5}}{\sqrt{5} +
\sqrt{6}} \omega_{Mie}. \label{classical}
\end{eqnarray}
Here $\omega_{Mie}$ is the Mie resonance frequency given by
\begin{equation}
\omega_{Mie} = \sqrt{\frac{1}{r_{s}^{3}}} \label{mie}
\end{equation}
and it is equal to $1/\sqrt{3}$ bulk plasmon frequency. The
classical expressions for $C^{AA}_{6}/R_{0}^{6}$ and
$C^{AA}_{8}/R_{0}^{8}$ given above are derived in Ref.
\cite{pacheco2} by assuming that all the strength of respective
multipole resonance is concentrated in a single peak.  By
substituting the values of $r_{s}$ in Eq. (\ref{classical}) we get
following numbers for sodium clusters $C^{AA}_{6}/R_{0}^{6}=
0.094$ and $C^{AA}_{8}/R_{0}^{8}= 0.49$ whereas for potassium
clusters $C^{AA}_{6}/R_{0}^{6}= 0.07$ and $C^{AA}_{8}/R_{0}^{8}=
0.37$. These classical numbers for $C^{AA}_{6}/R_{0}^{6}$ and
$C^{AA}_{8}/R_{0}^{8}$ are shown by asymptotic straight lines in
Figs. 1 and 2. We see from Figs. 1 and 2 that the
hydrodynamic approach yields correct asymptotic values of the van
der Waals coefficients or the values to which  these coefficient
saturate as the size of cluster grows.

To conclude, we have extended the applicability of the
hydrodynamic approach within TDDFT to the calculation of higher
order van der Waals coefficient $C^{AB}_{8}$ between clusters of
various sizes and different species. Our results for both
$C^{AB}_{6}$ and $C^{AB}_{8}$ are quite close to more accurate
orbital based TDKS approach. In particular we have found that the
numbers obtained by hydrodynamic approach for $C^{AB}_{8}$ are
more accurate than that of $C^{AB}_{6}$. For both sodium and
potassium clusters we have been able to calculate $C^{AB}_{6}$ and
$C^{AB}_{8}$ coefficients for clusters containing up to 1000
atoms. Thus we have been able to get the evolution of these
coefficients as a function of cluster size and have shown that
they approach their respective classical values for large
clusters.

\newpage
\begin{table}
\caption{Dispersion coefficient $C_{6}$ for sodium atom clusters
in atomic units (a.u.). The numbers follow the notation $3.60 (3)
= 3.60\times 10^{3}$. The numbers in parenthesis are results of
Refs. \cite{pacheco1,pacheco2} } \tabcolsep=0.1in
\begin{center}
\begin{tabular}{ccccc}
N & 2 & 8 & 20 & 40 \\
\hline
2 & 2.56(3) & 9.47(3) & 2.27(4) & 4.42(4)  \\
   & (2.62(3)) & (1.02(4)) & (2.45(4)) & (4.74 (4)) \\
8 &  & 3.51(4) & 8.42(4) & 1.64 (5)   \\
   &  & (4.01(4))& (9.55(4)) & ( 1.86 (5))  \\
20 &  &        & 2.02(5)& 3.93 (5)) \\
   &  &         & $(2.29(5))$ & ( 4.45 (5)) \\
40  &  &       &        & 7.63 (5)  \\
    &  &       &        & (8.60 (5)   \\
\end{tabular}
\end{center}
\end{table}

\begin{table}
\caption{Dispersion coefficient $C_{8}^{AB}$ for sodium atom
clusters in atomic units (a.u.). The numbers follow the notation
$3.60 (3) = 3.60\times 10^{3}$. The numbers in parenthesis are
results of Refs. \cite{pacheco1,pacheco2}} \tabcolsep=0.1in
\begin{center}
\begin{tabular}{ccccc}
N & 2 & 8 & 20 & 40 \\
\hline
2 & 3.08 (5) & 2.28 (6) & 8.95 (6)& 2.58 (7)  \\
   & (2.97(5)) & (2.27(6)) & (9.08(6)) & (2.69 (7)) \\
8 &  & 1.26(7) & 4.33 (7) & 1.15 (8)  \\
   &  & (1.32(7))& (4.59 (7)) & (1.25 (8))  \\
20 &  &        & 1.35 (8) & 3.37 (8)\\
   &  &         & (1.44 (8))& (3.67 (8))   \\
40 &  &          &        & 8.00 (8) \\
   &  &           &        & (8.82 (8)) \\
\end{tabular}
\end{center}
\end{table}
\begin{table}
\caption{Dispersion coefficient $C_{6}^{AB}$ for potassium atom
clusters in atomic units (a.u.). The numbers follow the notation
$3.60 (3) = 3.60\times 10^{3}$. The numbers in parenthesis are
results of Ref.\cite{pacheco2}} \tabcolsep=0.1in
\begin{center}
\begin{tabular}{cccc}
N& 2 & 8 & 20  \\
\hline
2 & 5.35 (3) & 2.03 (4) & 4.93 (4)\\
   & (6.28 (3)) & (2.34(4)) & (5.48(4)) \\
8 &  & 7.70 (4) & 1.87 (5)  \\
   &  & (8.71 (4)& (2.02 (5)) \\
20 &  &        & 4.55 (5) \\
   &  &         & (4.74 (5))  \\
\end{tabular}
\end{center}
\end{table}

\begin{table}
\caption{Dispersion coefficient $C_{8}^{AB}$ for potassium atom
clusters in atomic units (a.u.). The numbers follow the notation
$3.60 (3) = 3.60\times 10^{3}$. The numbers in parenthesis are
results of Ref. \cite{pacheco2} } \tabcolsep=0.1in
\begin{center}
\begin{tabular}{cccc}
N & 2 & 8 & 20  \\
\hline
2 & 8.69 (5) & 6.87 (6) & 2.77 (7)\\
   & (1.10 (6)) & (7.81 (6)) & (3.03 (7)) \\
8 &  & 3.97 (7) & 1.38 (8)  \\
   &  & (4.37 (7))& (1.43 (8)) \\
20 &  &        & 4.36 (8) \\
   &  &         & (4.55 (8))  \\
\end{tabular}
\end{center}
\end{table}

\begin{table}
\caption{Dispersion coefficient $C_{6}^{AB}$ for the pair
interaction between sodium clusters (values of N along the column)
and  potassium clusters (values of N along row) in atomic units
(a.u.). The numbers follow the notation $3.60 (3) = 3.60\times
10^{3}$. The numbers in parenthesis are results of
Ref.\cite{pacheco2}} \tabcolsep=0.1in
\begin{center}
\begin{tabular}{cccc}
N & 2 & 8 & 20  \\
\hline
2 & 3.68 (3) & 1.40 (4) & 3.40 (4)\\
   & (4.01 (3)) & (1.50 (4)) & (3.49 (4)) \\
8 & 1.35 (4) & 5.16 (4) & 1.26 (5)  \\
   & (2.16 (4)) & (8.05 (4)) & (1.88 (5)) \\
20 & 3.25 (4) & 1.24 (5)& 3.00 (5) \\
   &  (3.75 (4)) & (1.40 (5))& (3.26 (5))  \\
\end{tabular}
\end{center}
\end{table}

\begin{table}
\caption{Dispersion coefficient $C_{8}^{AB}$ for the pair
interaction between sodium clusters (values of N along the column)
and  potassium clusters (values of N along row) in atomic units
(a.u.). The numbers follow the notation $3.60 (3) = 3.60\times
10^{3}$. The numbers in parenthesis are results of Ref.
\cite{pacheco2}} \tabcolsep=0.1in
\begin{center}
\begin{tabular}{cccc}
N & 2 & 8 & 20  \\
\hline
2 & 5.21 (5) & 4.44 (6) & 1.84 (7)\\
   & (5.77 (5)) & (4.41 (6)) & (1.84 (7)) \\
8 & 3.55 (6) & 2.23 (7) & 8.29 (7)  \\
   & (4.52 (6)) & (2.91 (7)) & (1.12 (8)) \\
20 & 1.35 (7) & 7.30 (7)& 2.45 (8) \\
   &  (1.51 (7)) & (7.75 (7))& (2.57 (8))  \\
\end{tabular}
\end{center}
\end{table}

\clearpage
\newpage
\section*{Figure captions}
{\bf Fig.1}Plot of van der Waals coefficient $C^{AA}_{6}$ in units
of $R_{0}^{6}$ of alkali-metal clusters: sodium (solid circles)
and potassium (solid squares) as a function of $R_{0}$. The lines
are drawn as a guide to eye and horizontal lines represent
corresponding classical values of van der Waals coefficient.

{\bf Fig.2}Plot of van der Waals coefficient $C^{AA}_{8}$ in units
of $R_{0}^{8}$ of alkali-metal clusters: sodium (solid circles)
and potassium (solid squares) as a function of $R_{0}$. The lines
are drawn as a guide to eye and horizontal lines represent
corresponding classical values of van der Waals coefficient.
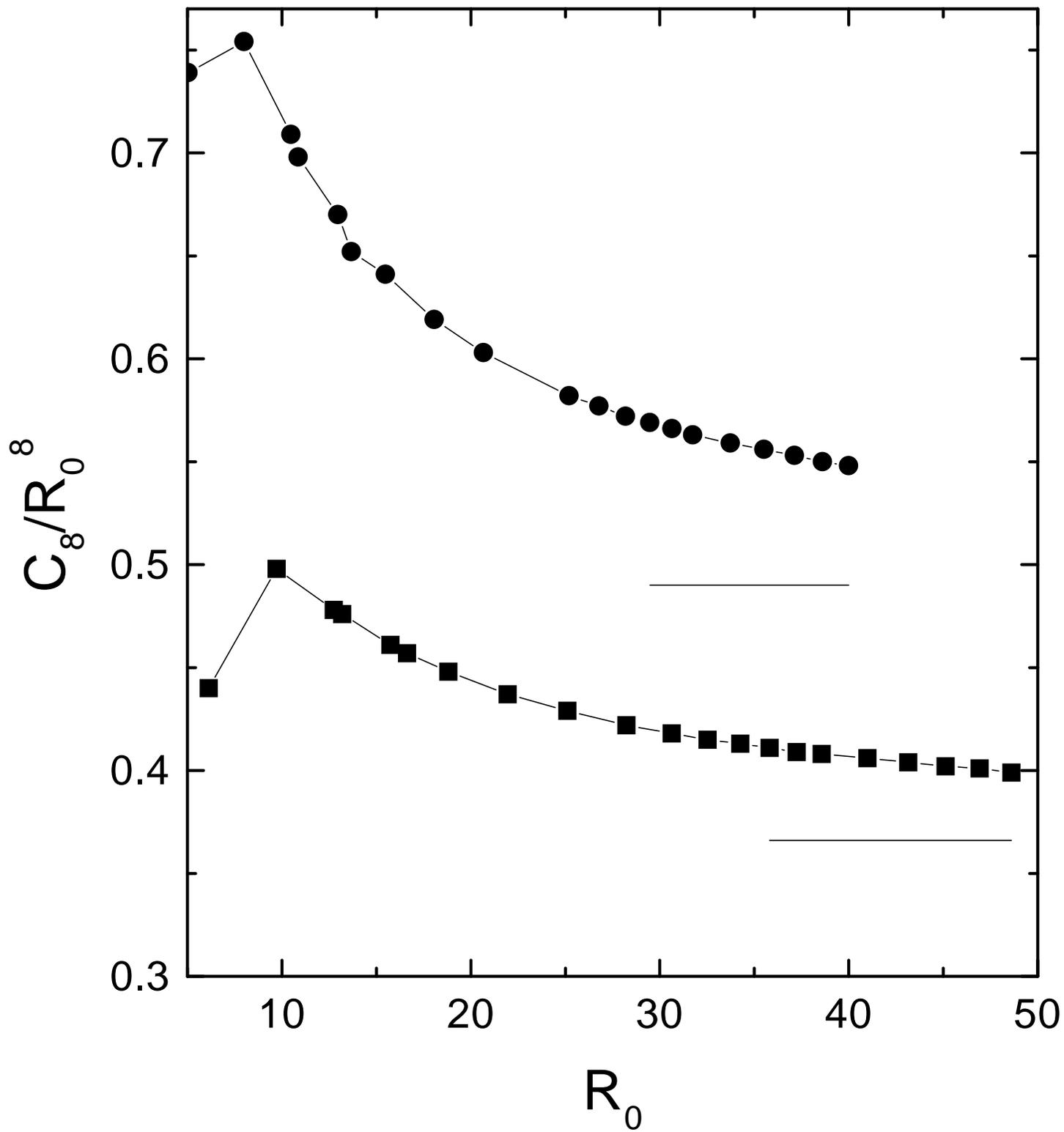
\end{document}